\documentclass[11pt]{article}
\usepackage{fullpage}

\usepackage{amsmath,amssymb}
\usepackage{hyperref}
\usepackage{graphicx}
\usepackage{latexsym}

\usepackage[utf8]{inputenc}
\DeclareUnicodeCharacter{00A0}{~}

\begin{document}

%%%%%%%%%%%%%%%%%%%%%%%%%%%%%%%%%%%%%%%%%%%%%%%%%%%%%%%%%
%
%  definitions

% i.e. and e.g.
\newcommand{\ie}{i.e.,\ }
\newcommand{\eg}{e.g.,\ }

% const
\newcommand{\const}{\operatorname{const.}} 

% sgn
%\newcommand{\sgn}{\operatorname{sgn}} 

% differentials (roman d)
\newcommand{\rmd}{\,\mathrm{d}}

% trace
\newcommand{\Tr}{\operatorname{tr}}

% Re and Im
\newcommand{\re}{\operatorname{Re}}
\newcommand{\im}{\operatorname{Im}}

% base of exponentials (roman e), with argument 
\newcommand{\e}[1]{\operatorname{e}^{#1}}

% life-time of black hole
\newcommand{\tBH}{t_{\text{ev}}}

% p_ort and p_par
\newcommand{\port}{{p_\perp}}
\newcommand{\ppar}{{p_\parallel}}

% diag
\newcommand{\diag}{\operatorname{diag}}

% order
\newcommand{\order}[1]{\operatorname{\mathcal{O}}\left(#1\right)}

%%%%%%%%%%%%%%%%%%%%%%%%%%%%%%%%%%%%%%%%%%%%%%%%%%%%%%%%%
%
%  title
 
\begin{center}

{\Large \textbf{Ladder operators for Klein-Gordon equation with scalar curvature term}}\\[2em]

\renewcommand{\thefootnote}{\fnsymbol{footnote}}
Wolfgang M{\"u}ck${}^{a,b}$\footnote[1]{E-mail: mueck@na.infn.it}\\[1em]%
\renewcommand{\thefootnote}{\arabic{footnote}}
${}^a$\emph{Dipartimento di Fisica ``Ettore Pancini'', Universit\`a degli Studi di Napoli "Federico II"\\ Via Cintia, 80126 Napoli, Italy}\\
${}^b$\emph{Istituto Nazionale di Fisica Nucleare, Sezione di Napoli\\ Via Cintia, 80126 Napoli, Italy}\\[2em]

\abstract{Recently, Cardoso, Houri and Kimura constructed generalized ladder operators for massive Klein-Gordon scalar fields in space-times with conformal symmetry. Their construction requires a closed conformal Killing vector, which is also an eigenvector of the Ricci tensor. 
Here, a similar procedure is used to construct generalized ladder operators for the Klein-Gordon equation with a scalar curvature term. It is proven that a ladder operator requires the existence of a conformal Killing vector, which must satisfy an additional property. This property is necessary and sufficient for the construction of a ladder operator. 
For maximally symmetric space-times, the results are equivalent to those of Cardoso, Houri and Kimura.}
\end{center}
\vspace{1em}

%%%%%%%%%%%%%%%%%%%%%%%%%%%%%%%%%%%%%%%%%%%%%%%%%%%%%%%%%
%
%  intro

\section{Introduction}
In two recent papers \cite{Cardoso:2017qmj, Cardoso:2017egd}, Cardoso, Houri and Kimura constructed ladder operators for the Klein--Gordan equation in manifolds possessing closed conformal Killing vectors, which are, in addition, eigenvectors of the Ricci tensor. 
More precisely, they constructed a first order operator $\mathcal{D}$ such that, if $\Phi$ is a solution of 
\begin{equation}
\label{KGE} 
	\left( \Box -m^2 \right) \Phi=0~,
\end{equation}
then $\mathcal{D}\Phi$ satisfies
\begin{equation}
\label{KGE2} 
	\left( \Box -m^2 -\delta m^2\right) \mathcal{D}\Phi=0~.
\end{equation}

Most of their examples involve maximally symmetric space-times, where the above conditions are satisfied. Because maximally symmetric space-times have constant curvature, one may wonder whether the mass term in \eqref{KGE} could be replaced by a scalar curvature term, which is quite natural from a geometrical point of view. 

Here, I will consider the Klein-Gordon equation 
\begin{equation}
\label{KGE.R}
	\left( \Box + \chi R \right) \Phi=0~,
\end{equation}
where $\Box=g^{\mu\nu} \nabla_\mu \nabla_\nu$ is the d'Alembertian (or Laplacian for a Riemannian manifold) and $\chi$ is a constant, which I will call the ``eigenvalue'', with some abuse of nomenclature. 
The scalar curvature $R$ is assumed to be non-vanishing. 
I will investigate, under which conditions there exists a first order ladder operator $\mathcal{D}$ that maps a solution of \eqref{KGE.R} to a solution of
\begin{equation}
\label{KGE.R2}
	\left( \Box + \chi' R \right) \mathcal{D} \Phi=0~.
\end{equation}
If it exists, the ladder operator $\mathcal{D}$ and the new eigenvalue will be determined.

%%%%%%%%%%%%%%%%%%%%%%%%%%%%%%%%%%%%%%%%%%%%%%%%%%%%%%%%%
%
%  main section

\section{Ladder operators from conformal Killing vectors}

\subsection{Properties of conformal Killing vectors}

Because conformal Killing vectors (CKVs) will play a crucial role in the construction of the ladder operator $\mathcal{D}$, I will start by recalling some of their basic properties. 

Consider a Riemannian or pseudo-Riemannian manifold of dimension $n$ admitting a CKV $\zeta$, 
\begin{equation}
\label{ckv.def}
	\nabla_\mu \zeta_\nu + \nabla_\nu \zeta_\mu = 2 Q g_{\mu\nu}~,
	\quad Q = \frac1n \nabla_\mu \zeta^\mu~. 
\end{equation}
Several identities derive from \eqref{ckv.def}. It is straightforward to obtain 
\begin{equation}
\label{ckv.id1} 
	\Box \zeta^\mu = -(n-2)\nabla^\mu Q - R^{\mu\nu}\zeta_\nu~.
\end{equation}
Differentiating this once more, one gets 
\begin{equation}
\label{ckv.id2} 
	\nabla_\mu \Box \zeta^\mu = -(n-2)\Box Q 
	- \frac12 \zeta^\mu \nabla_\mu R - RQ~.
\end{equation}
However, the left hand side of \eqref{ckv.id2} can also be written as 
\begin{equation}
\label{ckv.id3} 
	\nabla_\mu \Box \zeta^\mu = 
	\left[\nabla_\mu, \Box \right] \zeta^\mu 
	+ n \Box Q
	= \frac12 \zeta^\mu \nabla_\mu R  + RQ + n \Box Q~.
\end{equation}
Comparing \eqref{ckv.id2} and \eqref{ckv.id3}, one obtains the identity
\begin{equation}
\label{ckv.Q.id}
	\Box Q = \frac1{1-n} \left( RQ + \frac12 \zeta^\mu \nabla_\mu R \right)~.
\end{equation}
When acting on a scalar, the following commutation relation holds,
\begin{equation}
\label{D.comm}
	\left[ \Box , \zeta^\mu \nabla_\mu \right] = 2 Q\Box - (n-2) (\nabla^\mu Q) \nabla_\mu~.
\end{equation} 

\subsection{Equations for ladder operators}

Consider the first order operator 
\begin{equation}
\label{D1}
	\mathcal{D} = \eta^\mu \nabla_\mu + V~,
\end{equation}
where $\eta$ and $V$ are some vector and scalar, respectively. If $\mathcal{D}$ is a ladder operator in the sense of eqs.\ \eqref{KGE.R} and \eqref{KGE.R2}, then there must exist another first operator $\mathcal{D}' = \eta'{}^\mu \nabla_\mu +V'$ such that
\begin{equation}
\label{to.show}
	\left( \Box +\chi' R\right) \mathcal{D}
	- \mathcal{D}' \left( \Box +\chi R\right) =0~.
\end{equation}
Hence, the problem to be solved is to establish under which conditions one can find $\mathcal{D}$, $\mathcal{D}'$ and the new eigenvalue $\chi'$, given the eigenvalue $\chi$.  

By direct calculation, one finds
\begin{align}
\label{D.op}
	\left( \Box +\chi' R\right) \mathcal{D}
	- \mathcal{D}' \left( \Box +\chi R\right) 
	&= \left[\eta^\mu - \eta'{}^\mu\right] \nabla_\mu \Box 
	+ \left[ 2 (\nabla^\nu \eta^\mu) \nabla_\nu \nabla_\mu + (V-V')\Box \right] \\
\notag
	&\quad
	+ \left[ (\chi'\eta^\mu-\chi\eta'{}^\mu) R + 2\nabla^\mu V + R^\mu{}_\nu \eta^\nu + \Box \eta^\mu \right] \nabla_\mu \\
\notag 
	&\quad
	+ \left[ \Box V + (\chi' V- \chi V') R -\chi \eta'{}^\mu \nabla_\mu R\right]~.
\end{align}
To satisfy \eqref{to.show}, the terms collected in brackets on the right hand side of \eqref{D.op} must vanish separately. The term in front of the third order derivative simply yields
\begin{equation}
\label{eta.prime}
	\eta'{}^\mu =\eta^\mu~.
\end{equation}
The second order term vanishes, if and only if $\eta$ is a CKV, $\eta^\mu=\zeta^\mu$, and 
\begin{equation}
\label{V.prime}
	V' = V +2 Q~,
\end{equation}  
where $Q$ was defined in \eqref{ckv.def}. Thus, \eqref{eta.prime} and \eqref{V.prime} determine $\mathcal{D}'$, if $\mathcal{D}$ can be found.  

Using \eqref{eta.prime} and \eqref{V.prime} as well as the identies \eqref{ckv.id1} and \eqref{ckv.Q.id}, the two terms on the second and third lines of \eqref{D.op} give rise to the following two equations,
\begin{align}
\label{first.ord}
	(\chi' -\chi) R \zeta^\mu + 2 \nabla^\mu V - (n-2) \nabla^\mu Q &= 0~,\\
\label{zero.ord}
	(\chi' -\chi) R V + \Box V + 2 \chi (n-1) \Box Q &=0~.
\end{align}
Obviously, these always allow for the trivial solution $Q=V=0$, $\chi'=\chi$, in which $\zeta^\mu$ is a Killing vector. In this trivial case, the operator $\mathcal{D}$ is a symmetry operator.  
Henceforth, we shall assume non-zero $Q$.

Taking the divergence of \eqref{first.ord} and using again \eqref{ckv.Q.id}, one finds
\begin{equation}
\label{first.ord.div}
	2 \Box V - \left[n-2 +2 (n-1)(\chi'-\chi) \right] \Box Q + (\chi' - \chi)(n-2) RQ =0~.
\end{equation}
To proceed, let us introduce
\begin{equation}
\label{V.tilde}
	\tilde{V} = V +\gamma Q~,
\end{equation}
where $\gamma$ is a constant. A short calculation shows that, if $\gamma$ is chosen such that 
\begin{equation}
\label{gamma.def}
	\chi' -\chi = \frac{(n-2)\chi}{\gamma} - \frac{n-2+ \gamma}{n-1}~,
\end{equation} 
then \eqref{zero.ord} and \eqref{first.ord.div} can be combined into an equation involving only $\tilde{V}$,
\begin{equation}
\label{V.tilde.eq}
	\left[ (n-2+2\gamma) \Box + (n-2) (\chi'-\chi)R \right] \tilde{V} =0~.
\end{equation}
Note that $\gamma$ should be considered as a parameter, from which the new eigenvalue $\chi'$ is determined via \eqref{gamma.def}. To proceed further, one must distinguish the cases $n\neq 2$ and $n=2$.

\subsection{Case $n \neq 2$}

With some hindsight, introduce a new constant $\alpha$ by 
\begin{equation}
\label{alpha.def} 
	\chi' -\chi = -\frac{(n-2+2\gamma) \alpha}{n-1}~,
\end{equation} 
and define, for the sake of brevity,
\begin{equation}
\label{W.def}
	W = \frac{2\tilde{V}}{n-2+2\gamma}~.
\end{equation}
With \eqref{V.tilde}, \eqref{alpha.def} and \eqref{W.def}, equations \eqref{first.ord} and \eqref{V.tilde.eq} take the form 
\begin{align}
\label{A}
	\left( \nabla^\mu Q + \frac{\alpha}{n-1} R \zeta^\mu \right) &= \nabla^\mu W~,\\
\label{B}
	\left( \Box - \frac{n-2}{n-1} \alpha R \right) W &=0~,	
\end{align}
respectively. Using \eqref{gamma.def} and \eqref{alpha.def}, $\gamma$ is determined in terms of $\alpha$ and $\chi$ as one of
\begin{equation}
\label{gamma.alpha}
	\gamma = - \frac{(n-2)(1-\alpha)}{2(1-2\alpha)} 
		\left[\, 1 \pm \sqrt{ 1 + \frac{4(n-1)(1-2\alpha)\chi}{(n-2)(1-\alpha)^2}}\, \right]~.
\end{equation} 
We will comment on the case $\alpha=1/2$, which seems to be special, in a moment.

At this point, we can make the following observation. If, for a CKV $\zeta^\mu$, $\nabla^\mu Q$ is proportional to $R\zeta^\mu$, \ie if 
\begin{equation}
\label{suff.cond}
	\nabla^\mu Q + \frac{\alpha}{n-1} R \zeta^\mu =0
\end{equation}
holds for some constant $\alpha$, then \eqref{A} and \eqref{B} can be trivially solved by $W=0$. This implies that, with $\gamma$ given by \eqref{gamma.alpha}, the scalar $V$ in the ladder operator \eqref{D1} is $V=-\gamma Q$, and the new eigenvalue $\chi'$ follows from \eqref{alpha.def}. Therefore, \eqref{suff.cond} is a sufficient condition for the existence of a ladder operator. In the following, we show that the property \eqref{suff.cond} of the CKV is also a necessary condition.

To show this, consider \eqref{A} and \eqref{B}. The divergence of \eqref{A} can be used to eliminate $\Box W$ from \eqref{B}, which yields
\begin{equation}
\label{W.sol}
	W = Q + \frac{(n-1)(1-2\alpha)}{(n-2)\alpha} \frac1R \Box Q~.
\end{equation}
In passing, we note that the case $\alpha=1/2$ cannot be a solution, because \eqref{W.sol} would imply $W=Q$, which is inconsistent with \eqref{A}.
Substituting \eqref{W.sol} back into \eqref{A} and taking again the divergence gives
\begin{equation}
\label{Q2.eq}
	\Box \frac1R \Box Q + \frac{2(n-2) \alpha^2}{(n-1)(1-2\alpha)} \Box Q 
	- \frac{(n-2)^2 \alpha^2}{(n-1)^2(1-2\alpha)} RQ =0~.
\end{equation} 
This can be rewritten as 
\begin{equation}
\label{Q2.factor}
	( \Box + a_1 R) \frac1R ( \Box + a_2 R) Q = 0~,
\end{equation}
where $a_{1,2}$ are given by
\begin{equation}
\label{a12}
	a_{1,2} = \frac{(n-2)\alpha}{(n-1)(1-2\alpha)} \left[ \alpha \pm (\alpha-1) \right]~. 
\end{equation}
Therefore, $Q$ must satisfy either
\begin{equation}
\label{Q.diffeq}
	\left[ \Box +\frac{(n-2)\alpha}{(n-1)(1-2\alpha)} R \right]Q=0 \qquad \text{or} \qquad 
	\left[ \Box -\frac{(n-2)\alpha}{(n-1)} R \right]Q=0~.
\end{equation}
In the first case, \eqref{W.sol} gives $W=0$, which is the solution discussed above. In the second case, one gets 
$W=2\alpha Q$, so that \eqref{A} becomes
\begin{equation}
\label{A2}
	\nabla^\mu Q + \frac{\alpha}{(n-1)(1-2\alpha)} R\zeta^\mu =0~,
\end{equation} 
which is again of the form \eqref{suff.cond}, with $\tilde{\alpha}=\alpha/(1-2\alpha)$. 
Hence, we have shown that the property \eqref{suff.cond} of the CKV is a necessary and sufficient condition for the existence of a ladder operator.

It is interesting to note that, by virtue of the identity \eqref{ckv.Q.id}, the condition \eqref{suff.cond} implies
\begin{equation}
\label{suff.Q}
	\zeta^\mu \nabla_\mu R = 2\beta RQ~,\qquad \left( \Box + \frac{1+\beta}{n-1} R \right) Q=0~, 
\end{equation}
where 
\begin{equation}
\label{beta.def}
	\beta = \frac{n\alpha-1}{1-2\alpha}~, \qquad \alpha = \frac{1+\beta}{n+2\beta}~.
\end{equation}
Finally, it may also be useful to express \eqref{gamma.alpha} in terms of $\beta$, 
\begin{equation}
\label{gamma.beta}
	\gamma = -\frac12 (n-1 +\beta) \pm \frac12 \sqrt{(n-1 +\beta)^2 + 4 (n-1)(n+2\beta)\chi}~.
\end{equation}

\subsection{Case $n=2$}

For $n=2$, there is no need to introduce $\alpha$, because \eqref{gamma.def} reduces to
\begin{equation}
\label{gamma.def.2d}
	\chi' -\chi = -\gamma~,
\end{equation}
and \eqref{first.ord} and \eqref{zero.ord} become, with \eqref{V.tilde},
\begin{align}
\label{first.ord.2d}
	-\gamma R \zeta^\mu + 2 \nabla^\mu \tilde{V} -2\gamma \nabla^\mu Q &= 0~,\\
\label{zero.ord.2d}
	-\gamma R \tilde{V} +\gamma^2 RQ + \Box \tilde{V} + (2 \chi -\gamma) \Box Q &=0~.
\end{align}
The divergence of \eqref{first.ord.2d} implies 
\begin{equation}
\label{V.tilde.box.2d}
	\Box \tilde{V}=0~,
\end{equation}
so that \eqref{zero.ord.2d} gives
\begin{equation}
\label{V.tilde.2d}
	\tilde{V} = \gamma Q +\frac{2\chi -\gamma}{\gamma} \frac1R \Box Q~.
\end{equation}
Proceeding as in the case $n\neq2$, one can show that \eqref{V.tilde.box.2d} and \eqref{V.tilde.2d} allow only the solutions
\begin{equation}
\label{Q.diffeq.2d}
	\Box  Q =0 \qquad \text{or} \qquad 
	\left( \Box + \frac{\gamma^2}{2\chi-\gamma} R \right)Q=0~.
\end{equation}
In the first case, \eqref{V.tilde.2d} gives $\tilde{V}=\gamma Q$, which means $V=0$ from \eqref{V.tilde}. This, in turn, implies $\gamma=0$ from \eqref{first.ord.2d}, \ie $\chi'=\chi$. Thus, if $\Box Q=0$, which is equivalent to the statement that $R\zeta^\mu$ must be divergence free, then $\mathcal{D}=\zeta^\mu \nabla_\mu$ maps a solution of \eqref{KGE.R} to another solution with the same eigenvalue.

The second case of \eqref{Q.diffeq.2d} gives $\tilde{V}=0$, so that \eqref{first.ord.2d} and \eqref{zero.ord.2d} give rise to the two conditions
\begin{equation}
\label{cond.2d}
	R \zeta^\mu  +2 \nabla^\mu Q = 0~, \qquad 
	\left( \Box + \frac{\gamma^2}{2\chi -\gamma} R \right) Q =0~.
\end{equation}
Notice that these conditions are independent of each other, because the vector condition is divergence free. The eigenvalue shift $\gamma$ is to be determined from the scalar condition, in the sense that, if $Q$ satisfies
\begin{equation}
\label{cond.2d.Q}
	\left[ \Box + (1+\beta) R \right] Q=0
\end{equation}
for some $\beta$, then $\gamma$ is one of
\begin{equation}
\label{gamma.2d}
	\gamma = -\frac12 (1+\beta) \pm \frac12 \sqrt{(1+\beta)^2 + 8 (1+\beta)\chi}~.
\end{equation}
In the last two equations, we have adopted the notation of the general case, c.f.\ \eqref{suff.Q} and \eqref{gamma.beta}.

%%%%%%%%%%%%%%%%%%%%%%%%%%%%%%%%%%%%%%%%%%%%%%%%%%%%%%%%%
%
%  examples

\section{Corollary and examples}

One can observe that, in all cases, in which a ladder operator exists, $Q$ itself satisfies a Klein-Gordon equation of the form \eqref{KGE.R},
\begin{equation}
\label{cor.Q}
	( \Box + \chi R) Q=0~,\qquad \chi = \frac{1+\beta}{n-1}~,
\end{equation}
where we have adopted the notation used in \eqref{suff.Q}. Therefore, one can apply the results to construct the function 
\begin{equation}
\label{cor.Phi}
	\Phi = \mathcal{D} Q = \zeta^\mu \nabla_\mu Q - \gamma Q^2~,
\end{equation}
where $\gamma$ is one of the solutions of \eqref{gamma.beta},
\begin{equation}
\label{cor:gamma}
	\gamma \in (1+\beta, -n-2\beta)~.
\end{equation}
For $\gamma=1+\beta$, a short calculation shows that $\chi'=0$. Therefore, if $\Phi$ given in \eqref{cor.Phi} is non-zero, then it must satisfy the massless Klein-Gordon equation.

The simplest examples of ladder operators are, of course, those of a maximally symmetric space-time. Maximally symmetric space-times have $R_{\mu\nu} = \frac1n R g_{\mu\nu}$ with constant Ricci scalar $R$. This implies $\beta=0$ from \eqref{suff.Q}. Moreover, they possess closed CKVs, for which $\nabla_{[\mu} \zeta_{\nu]}=0$. Using \eqref{ckv.id1}, one can easily show that
\begin{equation}
\label{cckv}
	\nabla^\mu Q + \frac{1}{n(n-1)}R \zeta^\mu =0~,
\end{equation}
which is \eqref{suff.cond} with $\alpha=1/n$. Taking the case of AdS$_n$ with unit radius, where $R=-n(n-1)$, and writing $-\chi R=m^2$, one recovers the results for AdS$_n$ of \cite{Cardoso:2017qmj}.

As a non-trivial example, consider a spacially flat FLRW universe in $n=4$ dimensions,
\begin{equation}
\label{FLRW}
	\rmd s^2 = -\rmd t^2 + a^2(t) \left( \rmd r^2 + r^2 \rmd \Omega^2 \right)~.
\end{equation}
The vector $\zeta = a \partial_t$ is a time-like (closed) CKV, with $Q=\dot{a}$, $\nabla Q= \ddot{a} \rmd t$. Moreover, the Ricci scalar is 
\begin{equation}
\label{FLRW.Ricci}
	R = 6 \left( \frac{\ddot{a}}{a} + \frac{\dot{a}^2}{a^2} \right)~. 
\end{equation}
Taking, with hindsight,
\begin{equation}
\label{a.choice}
	a(t) = t^{-1/\beta}~,
\end{equation}
one can easily verify that \eqref{suff.cond} holds with $\alpha$ given by \eqref{beta.def}. In this example, $\Phi=\mathcal{D}Q=0$.

%%%%%%%%%%%%%%%%%%%%%%%%%%%%%%%%%%%%%%%%%%%%%%%%%%%%%%%%%
%
%  conclusions

\section{Conclusions}

Ladder operators for the Klein-Gordon equation with a scalar curvature term have been considered. It has been shown that ladder operators require the existence of a CKV. Furthermore, ladder operators exist, if and only if the CKV satisfies an additional property. This property, for dimensions $n\neq2$, is simply that $R\zeta^\mu$ must be proportional to $\nabla^\mu Q$. For $n=2$, there are two cases, which have been discussed in detail. In all cases, the ladder operator has the form
\begin{equation}
\label{conc.D}
	\mathcal{D} = \zeta^\mu \nabla_\mu - \gamma Q~,
\end{equation}
with a constant $\gamma$ that depends on the eigenvalue $\chi$ and a geometrical parameter that is involved in the additional property of the CKV.

The construction of the ladder operators is similar to \cite{Cardoso:2017qmj}, but appears to be somewhat more general, because the assumptions that the CKV be closed and an eigenvector of the Ricci tensor are replaced by the single requirement \eqref{suff.cond} (for $n\neq 2$). This simplification can be attributed to the use of a scalar curvature instead of a mass term in the Klein-Gordon equation. 
The results of \cite{Cardoso:2017qmj} for maximally symmetric space-times have been recovered and a simple non-trivial example has been provided. It would be interesting to find more examples of CKVs satisfying \eqref{suff.cond}, \eg among those given in \cite{Maartens:1995, Maartens:1996}.

\section*{Acknowledgments}
This research is supported in part by the INFN, research initiative STEFI.

%\bibliographystyle{utphys}
%\bibliography{ladder}

\providecommand{\href}[2]{#2}\begingroup\raggedright\endgroup

\end{document}